\documentclass[conference]{IEEEtran}
\usepackage{hyperref}
\IEEEoverridecommandlockouts
\usepackage{physics}
\usepackage{cite}
\usepackage{amsmath,amssymb,amsfonts}
\usepackage{algorithmic}
\usepackage{graphicx}
\usepackage{textcomp}
\usepackage{xcolor}
\def\BibTeX{{\rm B\kern-.05em{\sc i\kern-.025em b}\kern-.08em
    T\kern-.1667em\lower.7ex\hbox{E}\kern-.125emX}}
\begin{document}

\title{Surface Code Error Correction with Crosstalk Noise}

\author{\IEEEauthorblockN{Zeyuan Zhou}
\IEEEauthorblockA{\textit{Department of Computer Science}\\
\textit{Yale University}\\
\textit{Yale Quantum Institute}\\
New Haven, CT, USA}
\and
\IEEEauthorblockN{Andrew Ji}
\IEEEauthorblockA{\textit{University of Chicago} \\
\textit{Laboratory Schools}\\
Chicago, IL, USA}
\and
\IEEEauthorblockN{Yongshan Ding}
\IEEEauthorblockA{\textit{Department of Computer Science}\\
\textit{Department of Applied Physics}\\
\textit{Yale University}\\
\textit{Yale Quantum Institute}\\
New Haven, CT, USA}}

\maketitle
\begin{abstract}
The design and performance analysis of quantum error correction (QEC) codes are often based on incoherent and independent noise models since it is easy to simulate. However, these models fail to capture realistic hardware noise sources, such as correlated errors (crosstalk), which can significantly impact QEC code performance, especially when they occur between data and ancillary qubits. In this paper, we systematically study various types of crosstalk noise and quantify their effects on surface codes through memory and stability experiments. Based on our findings, we introduce crosstalk-robust implementations of QEC via flag qubit designs and redundant stabilizer checks. We perform both numerical and analytical studies to demonstrate the efficacy of these strategies. In addition, we analyze logical crosstalk in an $[[n,k>1,d]]$ code block and establish analytical conditions under which physical crosstalk does not lead to logical crosstalk. Together, our analytical and numerical results shed light on designing QEC codes that are robust against hardware realistic crosstalk noise, paving the way for reliable experimental realization of fault-tolerant quantum computing.
\end{abstract}
\section{Introduction}
Quantum error correction (QEC) emerges as a crucial component to realize large-scale fault-tolerant quantum computing (FTQC) \cite{Gottesman1998FTQC, nielsen2010quantum}. QEC encodes quantum information into a larger Hilbert space, which allows for detection and correction via repetitive measurements of its stabilizers. Traditionally, QEC studies assume an independent and incoherent noise model for stabilizer design and threshold simulation \cite{Gottesman1998FTQC}. This model, however, falls short in capturing all error properties of a physical system. For example, correlated error, also known as crosstalk, is a detrimental realistic noise on hardware. 

Generally characterized as an unwanted interaction, crosstalk takes various forms in hardware architectures. For example, in superconducting and semiconductor platforms, qubits are susceptible to parasitic $ZZ$ interaction from always-on coupling between neighboring qubits~\cite{krantz2019quanteng, Ash-Saki2020crosstalk,kandala2021transmons,zhao2022transmon}. In trapped-ion systems, unwanted $XX$-type interactions are commonly seen in the implementation of multi-qubit gates, such as the Mølmer–Sørensen gate \cite{Jungsang2022crosstalk, ParradoRodriguez2021crosstalk, Wineland2008trappedion}. Neutral atom systems also experience crosstalk when atom blockades are moved together to implement entangling gates \cite{Levine2018rydberg, Urban2009rydberg}. While crosstalk suppression has been thoroughly studied from both the hardware-level (e.g. coupler and gate design \cite{Houck2019coupler, Plourde2020hybrid, Yang2020anharmonicity}) and the software-level (e.g. compiling/mapping \cite{Zheng2022optimization, Ding2020crosstalk, Murali2020crosstalk, zhang2020slackqapproachingqubit, Zhang2022inverse,Thomsen2021Opti, Seif2024DD}, pulse design \cite{DasSarma2018DD, Mckay2022Hamiltonian, Tripathi2022DD,Brown2024chromatic, Coote2024DD, Zhou2023Crosstalk}), the investigation of its impact on QEC \cite{ParradoRodriguez2021crosstalk, Beale2018decoheres, Bravyi2018coherent, Greenbaum2018coherent, liu2025performance, Huang2019coherent, debroy2020performance, Iverson2020coherence} is still limited to certain platform-specific crosstalk type and lacks general measures to address crosstalk when instantiating a QEC code. 

Here, we provide a comprehensive analysis of the robustness of surface code, one of the most practical and scalable QEC codes, towards various types of crosstalk. Specifically, we perform memory and stability experiments for each code under a ``depolarizing + crosstalk" noise model, from which we find the effect of crosstalk on both state preservation and fault-tolerant logical operation. Both crosstalk and depolarizing noise models are circuit-level and are inspired by state-of-the-art qubits calibration and device connectivity data. Specifically, we survey through four types of crosstalk on two dimensions: gate-based or always-on, and between data-data qubits or between data-ancilla qubits. Numerical results show that among all four types, gate-based crosstalk between data and ancilla qubits poses the most detrimental effect on both memory and FTQC. To address this issue, we propose two schemes: a flagged syndrome extraction design for surface code and a redundant stabilizer check method suitable for repetition code. Both measures enable effective detection of crosstalk and are hardware agnostic, which means gate optimization or engineering is not required. The schemes are also generalizable to more codes. We perform both numerical and analytical studies to quantify the efficacy of these strategies and identify the noise regime where those techniques work the best. 

Apart from physical crosstalk on $[[n,1,d]]$ codes, we also introduce for the first time ``logical crosstalk" in an $[[n, k>1, d]]$ code block. We show that physical crosstalk can be converted to logical correlated errors contingent on the definition of logical Pauli operators. We show how logical crosstalk could impact logical gates using $[[4,2,2]]$ and as an example \cite{Linke2017FTQED, Honciuc_Menendez2024FT832} and calculated the effective logical crosstalk error rate under different physical noise model. Besides, we develop a crosstalk conversion condition and show analytically that on a $2$D sparse connectivity, low-weight logical crosstalk will not happen with large $d$. Together, these results fill the gap in the studies of how crosstalk impacts QEC and provide novel approaches to enhance crosstalk robustness.
\section{Noise Model}
We simulate a circuit-level noise model that includes both \emph{depolarizing} and \emph{crosstalk} noise. In this model, \emph{depolarizing noise} serves as the baseline, meaning that after each logical gate or reset operation, an incoherent Pauli error is inserted with a probability controlled by a single parameter \(p\). Formally, the depolarizing noise channel is expressed as:
\begin{eqnarray}
    \mathcal{E}_{1}(\rho) &=& (1-p_1)\rho 
    + \frac{p_1}{3} \sum_{i} \sigma^i \rho\, \sigma^i,\\
    \mathcal{E}_{2}(\rho) &=& (1-p_2)\rho 
    + \frac{p_2}{15} \sum_{i,j } 
    \bigl[\sigma^i \otimes \sigma^j\bigr]\rho \bigl[\sigma^i \otimes \sigma^j\bigr],
\end{eqnarray}
where \(i, j \in \{I, X, Y, Z\}\) and $(i,j)\neq (I,I)$. We parametrize all error probabilities with the single parameter \(p\) and scale it according to different gate scenarios. When conducting experiments, we vary \(p\) to observe how the logical error rates respond. The specific scaling factors for \(p\) in different gates are inspired by hardware characterization data across multiple quantum computing platforms. Beyond depolarizing noise, we also define \emph{crosstalk} noise as a correlated \(N\)-body Pauli error channel:
\begin{equation}
    \mathcal{E}_{N}(\rho) 
    = (1 - p_X)\,\rho + p_X \,\hat{P}\,\rho\, \hat{P}, 
    \quad \hat{P} = \hat{\sigma}_i^{\otimes N},
\end{equation}
where \(i \in \{X,Y, Z\}\). In particular, we consider four classes of crosstalk (see Table~\ref{tab1}) that are highly relevant to existing quantum hardware platforms. For example, in superconducting tunable qubit architectures, \(ZZ\)-type crosstalk typically dominates during two-qubit gates. Meanwhile, in fixed-frequency transmon qubits, crosstalk often appears as always-on \(ZZ\) coupling between nearest neighbors (N.N.)~\cite{Ash-Saki2020crosstalk, Ku2020hybrid, Gambetta2012crosstalk}. Similarly, in trapped-ion systems, residual \(XX\)-type coupling exists when implementing two-qubit gates~\cite{Jungsang2022crosstalk, ParradoRodriguez2021crosstalk, Wineland2008trappedion}. We categorize these crosstalk interactions as either \emph{gate-based} or \emph{always-on}, and they can occur between data and ancilla qubits or among data qubits themselves. In a 2D grid layout for syndrome extraction, data and ancilla qubits are typically placed as nearest neighbors to reduce routing overhead, hence enabling such crosstalk pathways. Additionally, data qubits can experience further crosstalk from next-nearest-neighbor (N.N.N.) interactions or multi-qubit gates. Notably, N.N.N.\ \(ZZ\)-type coupling is common in superconducting systems~\cite{zhao2020switch, PRXQuantum.4.010314}, though its strength can be orders of magnitude lower than N.N.\ coupling. Multi-qubit interactions frequently occur in neutral atom platforms where atom blockades are brought together for multi-qubit gates~\cite{Levine2018rydberg, Urban2009rydberg}. For \emph{always-on crosstalk}, we parametrize the crosstalk error through a Pauli-twirling approximation (PTA) as follows:
\begin{eqnarray}
    p_{ZZ}(t) \;=\; \sin^2\bigl(J_{ZZ}\, t\bigr),
\end{eqnarray}
where the coefficient \(J_{ZZ}\) represents the coupling strength. Here, we use publicly available IBM Quantum device data to set \(J_{ZZ} = 10\)\,kHz~\cite{Tripathi2022DD, Zhou2023Crosstalk}, and use the gate durations specified in Table~\ref{tab:operation_durations} to determine the effective crosstalk probability. Next, we discuss the origin and details of the four crosstalk types that we incorporate into our experiments. 

\begin{enumerate}
\item \textit{Gate-Based Data--Ancilla Crosstalk}\\
This class of crosstalk occurs between data and ancilla qubits, which are often nearest neighbors in many quantum error-correcting codes. It is inserted only after two-qubit gates, reflecting the scenario in which crosstalk happens while qubits are actively interacting (e.g., via CNOT or CZ). Such gate-based crosstalk is commonly observed in tunable transmon or coupler architectures, including those used by Google Quantum.

\item \textit{Always-on Data--Ancilla Crosstalk}\\
Another type of data--ancilla interaction is always-on crosstalk. In this scenario, crosstalk is present regardless of whether qubits are implementing a gate or simply idling. We incorporate this effect by adding incoherent \(ZZ\) gates with probabilities scaled to the active or idle durations in Table~\ref{tab:operation_durations}. Such always-on crosstalk is frequently observed in fixed-frequency transmon qubits (e.g., IBM Quantum hardware).

\item \textit{Gate-Based Data--Data Crosstalk}\\
Next-nearest-neighbor (N.N.N.) crosstalk represents another major noise source, and we model it here as data--data crosstalk. Even if there are no direct operations between two N.N.N.\ data qubits, they can become resonant if the intermediate ancilla qubit is engaged in gate operations. This phenomenon is commonly seen in trapped-ion platforms. To simulate it, we apply gate-based crosstalk across all N.N.N.\ data--data pairs after each time step in the syndrome measurement (there are four time steps per round).

\item \textit{Always-on Data--Data Crosstalk}\\
Finally, this type of crosstalk also affects the same qubit pairs as in Type~III but remains active at all times, not just during gate operations. Consequently, we scale the error probabilities by the durations in Table~\ref{tab:operation_durations}, allowing crosstalk to persist continuously during both active gates and idle periods.
\end{enumerate}
\begin{table}[htbp]
\caption{Operation durations for different quantum operations and the device reference \(T_1\) and \(T_2\) times.}
\label{tab:operation_durations}
\centering
\begin{tabular}{l c}
\hline\hline
\textbf{Operations} & \boldmath$t_g (\mu s)$ \\
\hline
H, S gate       & 0.02 \\
CNOT, CZ gate   & 0.04 \\
Measurement     & 0.60 \\
Reset           & 0.50 \\
Reference $T_1$ & 30.0 \\
Reference $T_2$ & 30.0 \\
\hline\hline
\end{tabular}
\end{table}

\begin{figure}[htbp]
\centerline{\includegraphics[width=\columnwidth]{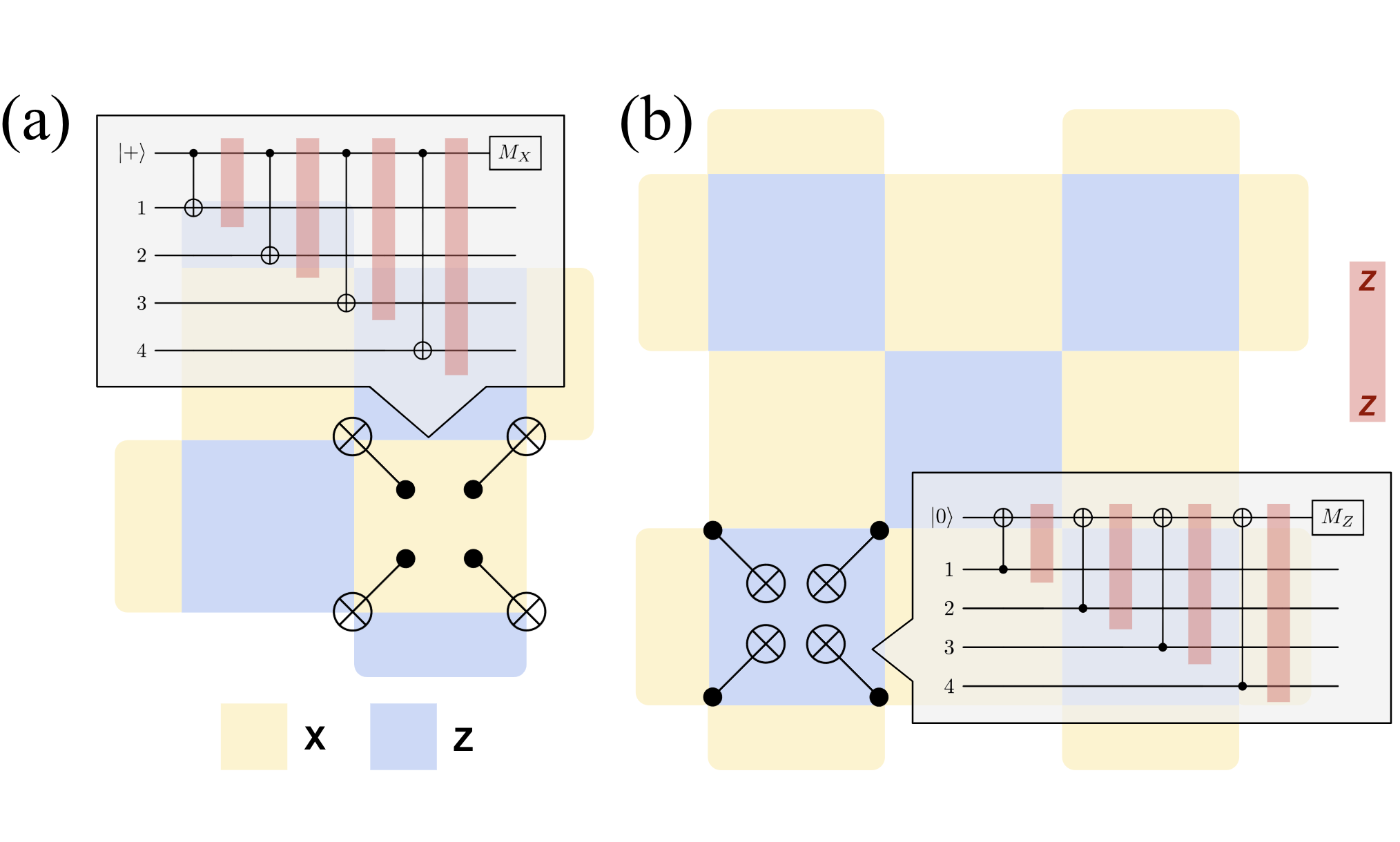}}
\caption{The rotated surface code patch (a) and the stability experiment patch (b). The yellow (blue) plaquettes 
are $X(Z)$ stabilizers with support on data qubits at the vertices, and the check qubits are placed in the middle. 
Red shaded area denotes $ZZ$ correlated crosstalk after each two-qubit gate in syndrome measurement circuits.}
\label{fig1}
\end{figure}

\subsection{Coherence vs Incoherence}
Always-on crosstalk is shown to be coherent in current quantum devices, which is hard to fully simulate. In this paper, we consider the Pauli-twirled incoherent simulation of such crosstalk. In the setting of a code-capacity model, a great number of studies have been conducted to study the coherence in error correction and how accurate the Pauli-twirled approximation (PTA) is in quantifying the effect of  coherent noise. Assuming noiseless syndrome measurement cycles, conclusions are reached from both analytical \cite{Greenbaum2018coherent, Beale2018decoheres} and numerical studies that the ratio between coherent and incoherent contributions will decay exponentially by scaling up code distance. This implies that an incoherent noise model makes a close approximation for high distances. In a circuit-level noise model, where error (including crosstalk) can occur during syndrome measurements, we analyze a simple case shown as below to demonstrate the discrepancy between incoherent and coherent channels.

Let's first consider an $X$-stabilizer check circuit shown in Fig \ref{fig1}. Ancilla qubit is initialized in $\ket{+} = \frac{1}{\sqrt{2}}(\ket{0}+\ket{1})$ state, and the four data qubits are in some unknown entangled state $\ket{\psi}_D = \sum_{x\in\{0,1\}^4}\alpha_x\ket{x_1x_2x_3x_4}$ ($\ket{x}=\ket{x_1x_2x_3x_4}$), the total initial state of the system is $\ket{\Psi}_0 = \frac{1}{\sqrt{2}}(\ket{0}+\ket{1})\ket{\psi}_D$. After four consecutive CNOTs, the state becomes $\ket{\Psi}_1 = \frac{1}{\sqrt{2}} \sum_{x} \alpha_x\big[\ket{0}\ket{x} + \ket{1}\ket{x\oplus 1^4} \big]$. Due to commutivity relations, all the correlated $ZZ(\theta)$ errors can be propagated to after CNOT gates. Here we applies to each (ancilla, data) pair $U_{ZZ}(\theta) = \exp\big[-i\frac{\theta_i}{2}Z_0Z_i\big]$, and the total unitary $U_{ZZ} = \prod_{i=1}^4 \exp\big[-i\frac{\theta_i}{2}Z_0Z_i \big]$. Since $Z_0Z_i\ket{p}_0\ket{q_i}_i = (-1)^{p+q_i}\ket{p}_0\ket{q_i}_i$, the combined phase after four correlated phase errors is $\exp\big[-\frac{i}{2}\sum_{i=1}^{4}\theta_i(-1)^{p+x_i}\big]$. The resulting state becomes: 
\begin{equation}
    \ket{\Psi_2} = \frac{1}{\sqrt{2}}\sum_{x}\alpha_x e^{i\phi(x)} \bigg[\ket{0}\ket{x} + \ket{1}\ket{x\oplus 1^4}\bigg],
\end{equation}
where $\phi(x) = -\frac{1}{2}\sum_{i=1}^{4}\theta_i(-1)^{x_i}$. Finally after the Hadamard gate on ancilla qubit, the state becomes: 
\begin{equation}
\ket{\Phi} = \frac{1}{2}\sum_x\alpha_x e^{i\phi(x)}\bigg[\ket{0}(\ket{x}+\ket{x\oplus 1^4} + \ket{1}(\ket{x} - \ket{x\oplus1^4})\bigg]
\end{equation}
Here, the ancilla qubit picks up phases determined by the actual state from $\phi(x)$; and the phases will destructively or constructively interfere to accumulate unless trivial $\theta_i$ value. 
The incoherent channel, however, is a classical mixture over $16$ possible subsets $S$, which can be expressed as: 
\begin{equation}
    \mathcal{E}_{\textrm{inc}}(\rho) = \sum_{S\in \{1,2,3,4\}}\bigg(\prod_{i\in S}p_i\bigg)\bigg(\prod_{j\notin S}[1-p_j]\bigg)(Z_0Z_S)\rho(Z_0Z_S),
\end{equation}
where $Z_0Z_S = \prod_{i\in S}Z_AZ_{Di}$, and we try to match each $\theta_i$ by setting $p_i = \sin^2(\theta_i/2)$. However, each of the $16$ possible situations are independent and cannot possibly mimic the full coherent scenario, causing a discrepancy of final state on the ancilla qubit due to phase accumulation. Similarly, for $Z$-stabilizer circuit, the four data qubits will accumulate phases from crosstalk in other qubit pair due to the propagation through CNOTs, but the ancilla qubits will be unaffected by the $ZZ$ errors since it is in the $Z$-basis. This analysis shows that coherence cannot be fully mimicked using incoherent noise model for nontrivial circuits under circuit level noise. It is still an open question of efficiently quantifying the impact of coherence on large-scale QEC codes. 

\begin{table*}[htbp]
\caption{Comparison of various crosstalk types and their impact on a rotated surface code. 
The first row shows the baseline depolarizing noise model without crosstalk, while all 
subsequent rows use the same noise model plus the corresponding crosstalk type on specified qubit pair. 
Parameters such as coupling strength and error rates are inspired by hardware data. Memory 
experiments yield the code threshold and effective distance $d_{\mathrm{eff}}$, 
where $(d)$ indicates the code distance. The time overhead ratio $R$ is extracted from 
stability experiments, with $(w)$ specifying the patch width. The last column identifies 
the error source.}
\label{tab1}
\centering
\renewcommand{\arraystretch}{1.2} 
\begin{tabular}{p{2.0cm} p{1.5cm} p{1.5cm} p{1.8cm} p{1.6cm} p{1.8cm} p{1.8cm} p{2.7cm}}
\hline\hline
\textbf{Crosstalk Type} & \textbf{Qubit Pair} & \textbf{Coupling Strength (GHz)} & 
\textbf{Error Rate} & \textbf{Threshold [\%]} & \boldmath$d_{\mathrm{eff}}(d)$ & 
\boldmath$R(w)$ & \textbf{Error Source} \\
\hline
(Baseline) Depolarizing error, no crosstalk
& single-qubit, two-qubit 
& 
& $p_{\mathrm{2Q}} = p$ \newline
  $p_{\mathrm{1Q}} = 0.1p$ \newline
  $p_{\mathrm{reset}} = 2p$ \newline
  $p_{\mathrm{meas}} = 5p$
& 0.74
& $1.94 \pm 0.02 (3)$ \newline
  $2.98 \pm 0.06 (5)$ \newline
  $3.98 \pm 0.06 (7)$ \newline
  $4.91 \pm 0.10 (9)$
&
&
Single-qubit gate error, two-qubit gate error, reset error, and measurement readout error 
\\

Gate-based
& Data-Ancilla
& 
& $p_{ZZ}=10^{-3}$
& 0.63
& $1.40 \pm 0.06 (3)$ \newline
  $2.07 \pm 0.09 (5)$ \newline
  $2.79 \pm 0.02 (7)$ \newline
  $3.57 \pm 0.01 (9)$
& $1.15 \pm 0.03 (4)$ \newline
  $1.14 \pm 0.02 (6)$ \newline
  $1.14 \pm 0.02 (8)$
& N.N. parasitic $ZZ$-type interaction during entangling gates 
\\

Always-on
& Data-Ancilla
& $J = 10^{-5}$
& 
& 0.71
& $1.83 \pm 0.01 (3)$ \newline
  $2.79 \pm 0.01 (5)$ \newline
  $3.93 \pm 0.08 (7)$ \newline
  $4.62 \pm 0.16 (9)$
& $1.00 \pm 0.02 (4)$ \newline
  $1.00 \pm 0.01 (6)$ \newline
  $1.00 \pm 0.01 (8)$
& N.N. coherent $ZZ$-rotation due to residual coupling 
\\

Gate-based
& Data-Data
& 
& $p_{ZZ} = 10^{-4}$
& 0.66
& $0.72 \pm 0.10 (3)$ \newline
  $1.57 \pm 0.12 (5)$ \newline
  $2.13 \pm 0.18 (7)$ \newline
  $2.86 \pm 0.20 (9)$
& $1.01 \pm 0.03 (4)$ \newline
  $0.97 \pm 0.01 (6)$ \newline
  $1.00 \pm 0.02 (8)$
& N.N.N. $ZZ$ interaction during simultaneous entangling gate due to unwanted resonance
\\

Always-on
& Data-Data
& $J = 10^{-5}$
& 
& 0.71
& $1.07 \pm 0.08 (3)$ \newline
  $2.11 \pm 0.12 (5)$ \newline
  $2.94 \pm 0.15 (7)$ \newline
  $3.65 \pm 0.26 (9)$
& $0.97 \pm 0.02 (4)$ \newline
  $1.02 \pm 0.02 (6)$ \newline
  $1.00 \pm 0.01 (8)$
& N.N.N. static $ZZ$-rotation due to $2$nd order residual coupling 
\\
\hline\hline
\end{tabular}
\end{table*}

\section{Code Performance}
\subsection{Stabilizer Codes}
We briefly introduce the stabilizer formalism for defining quantum error correction codes. An $[[n,k,d]]$ code encodes $k$ qubits of logical information with $n$ physical qubits to distance $d$, enabling correction of up to $\lfloor(d-1)/2\rfloor$ errors and detection of up to $(d-1)$ errors. The \textit{stabilizer group} $\mathcal{S}$ $\subseteq$ $\mathcal{P}_n$ (Pauli group on $n$ qubits) is generated by $\langle S_1, S_2, \ldots, S_{n-k} \rangle$ such that $[S_i, S_j] = 0, \forall i, j$, where each $S_i \in \mathcal{P}_n$ is of the form $ \bigotimes_{j=1}^{n}P_j$ with $P_j \in \{I, X, Y, Z\} $. Then we can define a complete set of orthogonal projectors $\Pi_{\vec{s}} = \prod_{i=1}^{n-k} \frac{1}{2} \left( I + (-1)^{s_i} S_i \right)$, where $s_i\in \{0,1\}$ is the syndrome outcome. Based on the syndrome information, a Pauli operator $R_{\vec{s}}$ can be found to map faulty state $\ket{\phi}$ back to correct codeword state $R_{\vec{s}}\Pi_{\vec{s}}\ket{\phi}$. 

Our primary focus is the rotated surface code, which has been well studied both theoretically and numerically \cite{nielsen2010quantum, Gottesman1998FTQC, Fowler2012surface, Horsman2012surgery}, and has been demonstrated on hardware platforms with exponential error suppression capabilities \cite{Chen2021repetition, google2023suppressing, Zhao2022surface}. The rotated surface code is a generalized toric code that can be implemented on a planar surface without open boundary conditions. Surface code allows correction of both $X$ and $Z$ errors by checking weight-$4$ local stabilizers. Due to its high threshold and large number of logical gates that can be implemented, rotated surface codes arise as one of the most promising candidates for FTQC. Rotated surface code encodes $k=1$ logical qubit using $n=d^2$ data qubits and $(d^2-1)$ ancilla qubits. In most surface code instantiations, data qubits are coupled to only ancilla qubits and vice versa. Fig. \ref{fig1} shows the $X$- and $Z$-stabilizer syndrome measurement circuits for the rotated surface code. 
\subsection{Memory Experiments}
\begin{figure}[htbp]
\centerline{\includegraphics[width=0.9\columnwidth]{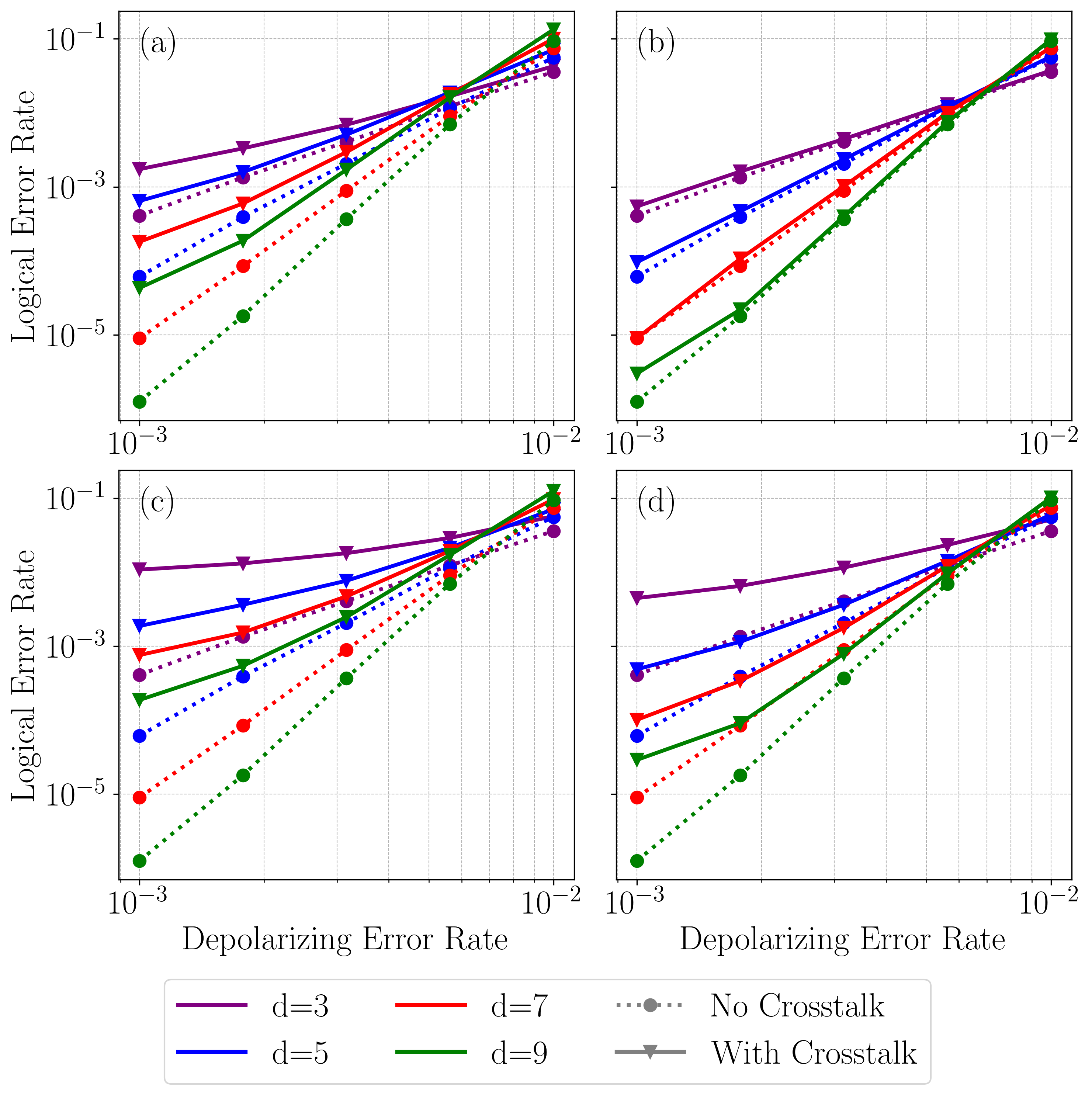}}
\caption{Comparison of different crosstalk effects on the logical error rate of the surface code $X$-memory. The plots show: 
        (a) Gate-based crosstalk between data and ancilla qubits with error probability \( p = 0.001 \). 
        (b) Always-on crosstalk between data and ancilla qubits with coupling strength \( J = 0.00001 \). 
        (c) Gate-based next-nearest-neighbor crosstalk between data qubits with \( p = 0.0001 \).
        (d) Always-on next-nearest-neighbor crosstalk between data qubits with \( J = 0.00001 \). 
        In all cases, dotted lines represent baseline performance without crosstalk, while solid lines show performance with crosstalk effects for different distances.}
\label{fig4}
\end{figure}

With the noise model described above, we simulated the memory of surface code patches with various distances to examine the impact of crosstalk on logical error rates. We simulated surface codes with distance $d_s\in \{3, 5,7,9\}$ and with repeated syndrome measurements of $d_s$ times for fault tolerance. For each circuit, we first initialize the surface code qubit to be in logical $\ket{+}$ state (which is susceptible to $ZZ$ crosstalk), then we scale the depolarizing noise parameter $p$ from $10^{-3}$ to $10^{-2}$. The crosstalk error rate for each type is extracted from hardware data and is held constant throughout the simulation. The error parameters are specified in Table \ref{tab1}, with gate-based crosstalk specified by an error rate following each two-qubit gate, and always-on crosstalk is parameterized with coupling strength $J$. The circuits are set up by the Stim package \cite{Gidney2021Stim} and decoded with the Pymatching package via Minimum Weight Perfect Matching (MWPM) algorithms \cite{higgott2021pymatching}. For each circuit, Monte Carlo simulation is conducted with $10^6$ shots. For baseline crosstalk-free noise and each type of crosstalk, we plot the depolarizing error rate $p$ versus logical error rate $p_L$ and extract the threshold value $(\%)$ and the effective distance $d_{\mathrm{eff}}$. Here, $d_{\mathrm{eff}}$ is extracted as the exponent of $p$ in the approximate
expression for $p_L$ as $p_L \propto p^{d_{\mathrm{eff}}}$, which can generally be understood as the minimum number of errors needed to create a logical error. The physical error rate versus logical error rate plots for all $4$ types of crosstalk is shown in Fig. \ref{fig4}, and the code threshold and effective distances are shown in Table \ref{tab1}. From the memory experiment, we observed that among all $4$ types, gate-based crosstalk between data and ancilla qubits poses the most amount of reduction in threshold value (from $0.74$ to $0.63$), and gate-based data-data crosstalk poses second detrimental effect on threshold (from $0.74$ to $0.66$). This implies that syndrome extraction steps are greatly degraded due to the correlated events created by such crosstalk, causing a stricter requirement for gate fidelity for fault tolerance. One the other hand, gate-based data-data crosstalk significantly shifts up the logical error rate and lower down $d_{\mathrm{eff}}$ due to the reason that it creates more two-qubit correlated error and hence more likely to cause a logical error. Static always-on crosstalk, either between data-ancilla or data-data, causes relatively smaller changes to both the threshold and effective distances. 

\subsection{Stability Experiments}
\begin{figure}[htbp]
\centerline{\includegraphics[width=0.85\columnwidth]{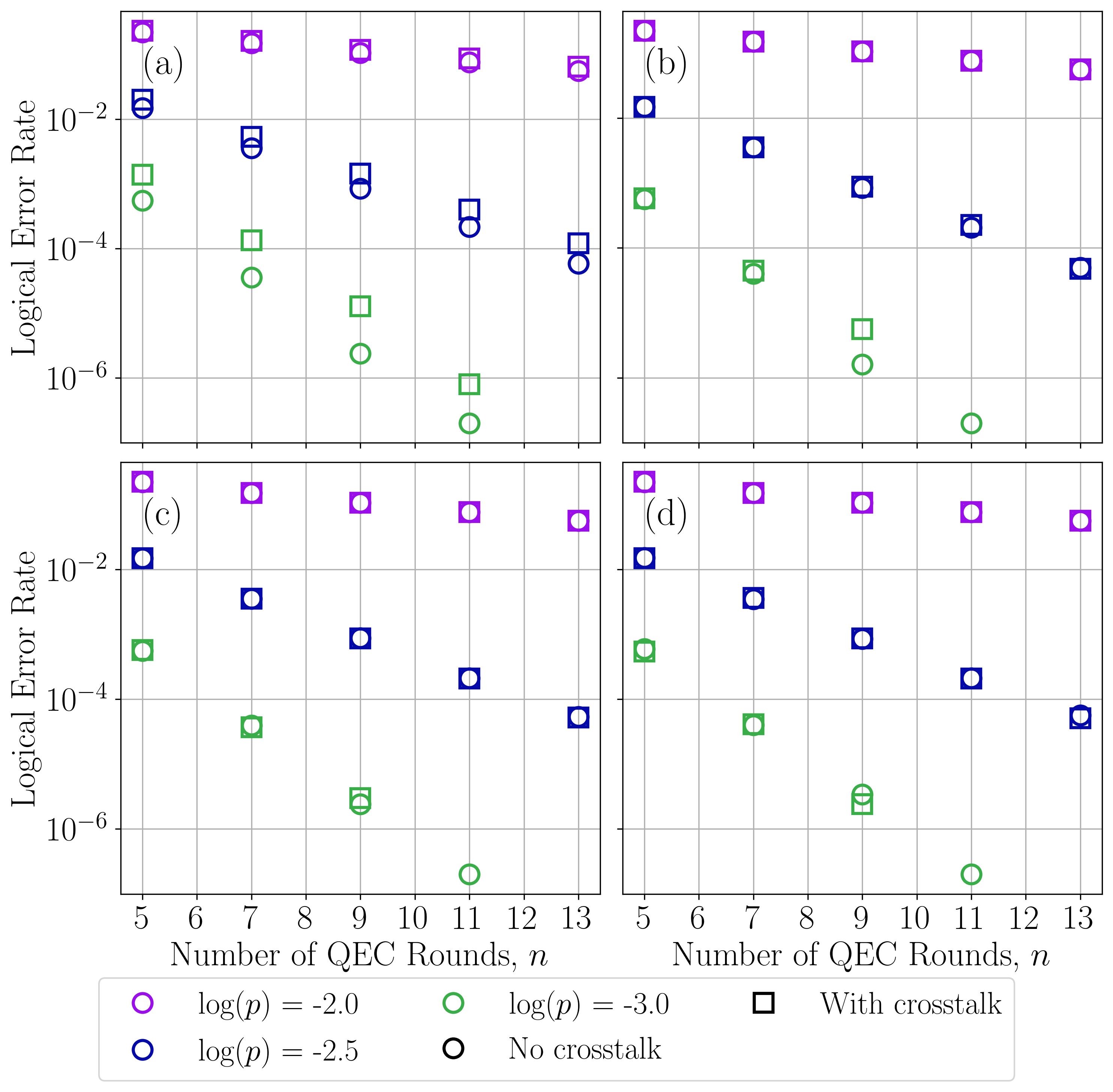}}
\caption{Comparison of different crosstalk effects on the logical error rate of the surface code stability experiments. The plots show: 
        (a) Gate-based crosstalk between data and ancilla qubits with error probability \( p = 0.001 \). 
        (b) Always-on crosstalk between data and ancilla qubits with coupling strength \( J = 0.00001 \). 
        (c) Gate-based next-nearest-neighbor crosstalk between data qubits with \( p = 0.0001 \).
        (d) Always-on next-nearest-neighbor crosstalk between data qubits with \( J = 0.00001 \). 
        In all cases, circles represent baseline performance without crosstalk, while squares show performance with crosstalk effects for different distances.}
\label{fig7}
\end{figure}

\begin{figure}[htbp]
\centerline{\includegraphics[width=0.9\columnwidth]{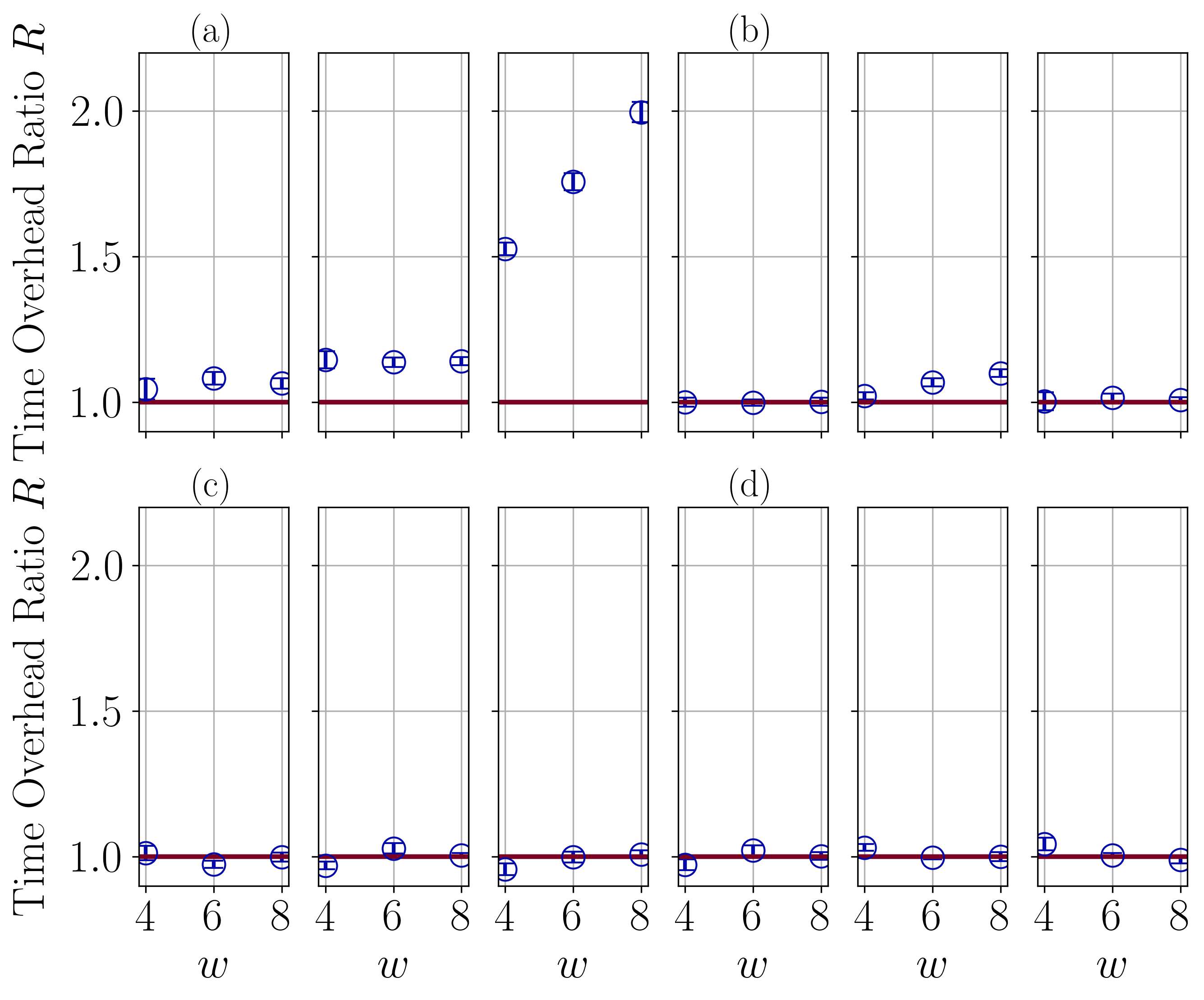}}
\caption{Comparison of different crosstalk effects on the Time Overhead Ratio $R$ of the surface code stability experiments with depolarizing noise $p=10^{-2.5}$. The plots show: 
        (a) Gate-based crosstalk between data and ancilla qubits. 
        (b) Always-on crosstalk between data and ancilla qubits. 
        (c) Gate-based next-nearest-neighbor crosstalk between data qubits.
        (d) Always-on next-nearest-neighbor crosstalk between data qubits. 
        The red line represents the break-even point. }
\label{fig5}
\end{figure}

While memory experiments probe how logical states persist through time, they do not address how errors affect the spatial propagation of logical observables, a key requirement for fault-tolerant quantum computing (FTQC). We use stability experiments to examine the impact of crosstalk noise on stabilizer measurements~\cite{Gidney2022stability,gehér2024resetresetquestion}, which are essential for determining joint logical Pauli products in lattice surgery~\cite{Horsman2012surgery, Litinski2019surgery,Chamberland2022surgery,Bombin2023FTQC}. Stability experiments verify a global invariant across space by ensuring that the product of stabilizer measurements equals a predefined value. Here, we consider $w \times w$ stability patches with $w \in \{4,6,8\}$ and a depolarizing noise parameter $p \in \{10^{-2}, 10^{-2.5}, 10^{-3}\}$. Although these patches do not encode a logical state, they include one type of predetermined stabilizer. In each run, all data qubits are prepared in the logical $\ket{+}$ state, the stabilizer is measured for $n$ rounds, and finally all data qubits are measured. We determine the logical error rate $p_L$ as a function of $n$, observing an exponential decay $p_L = A e^{-\gamma n}$, where $\gamma$ and $A$ are extracted from a linear fit of $\log(p_L)$ versus $n$. Fig.~\ref{fig7} plots the decay in $p_L$ versus the number of QEC rounds $n$. We conduct this analysis for circuits with ($\gamma_{\text{ct}}$) and without ($\gamma_0$) crosstalk noise. To quantify the crosstalk effect, we define the time overhead ratio $R = \gamma_0 / \gamma_{\text{ct}}$, indicating how many additional stabilizer measurement rounds are required, due to crosstalk, to reach the same logical error rate. For each crosstalk type, we also test $p = 10^{-2.5}$ under three crosstalk strengths: half, the same, of double of the strength of crosstalk specified in column $2$ and $3$ of Table \ref{tab1}. A larger ratio of $R > 1$ shows how greater the crosstalk has degraded the syndrome measurement accuracy. Fig.~\ref{fig5} plots the time overhead ratio of the $4$ types of crosstalk under three different crosstalk strength for each patch width $w$. Only gate-based data--ancilla crosstalk significantly increases both the logical error rate and $R$. Since stability experiments mainly detect time-like logical errors from repeated stabilizer measurement failures, higher ancilla error rates due to crosstalk raise the likelihood of logical failure, necessitating more stabilizer rounds to achieve a target error rate. Other types of crosstalk, however, don't pose any significant changes to either the logical error rate of the time overhead ratio. As a proxy for FTQC, these results imply an intuitive understanding that data-ancilla crosstalk remains the most detrimental in lattice surgery operations where logical operator measurement is the key component that determines the logical gate fidelity.  

\section{Crosstalk Robust QEC}
\begin{figure*}[htbp]
\centering
\includegraphics[width=0.8\textwidth]{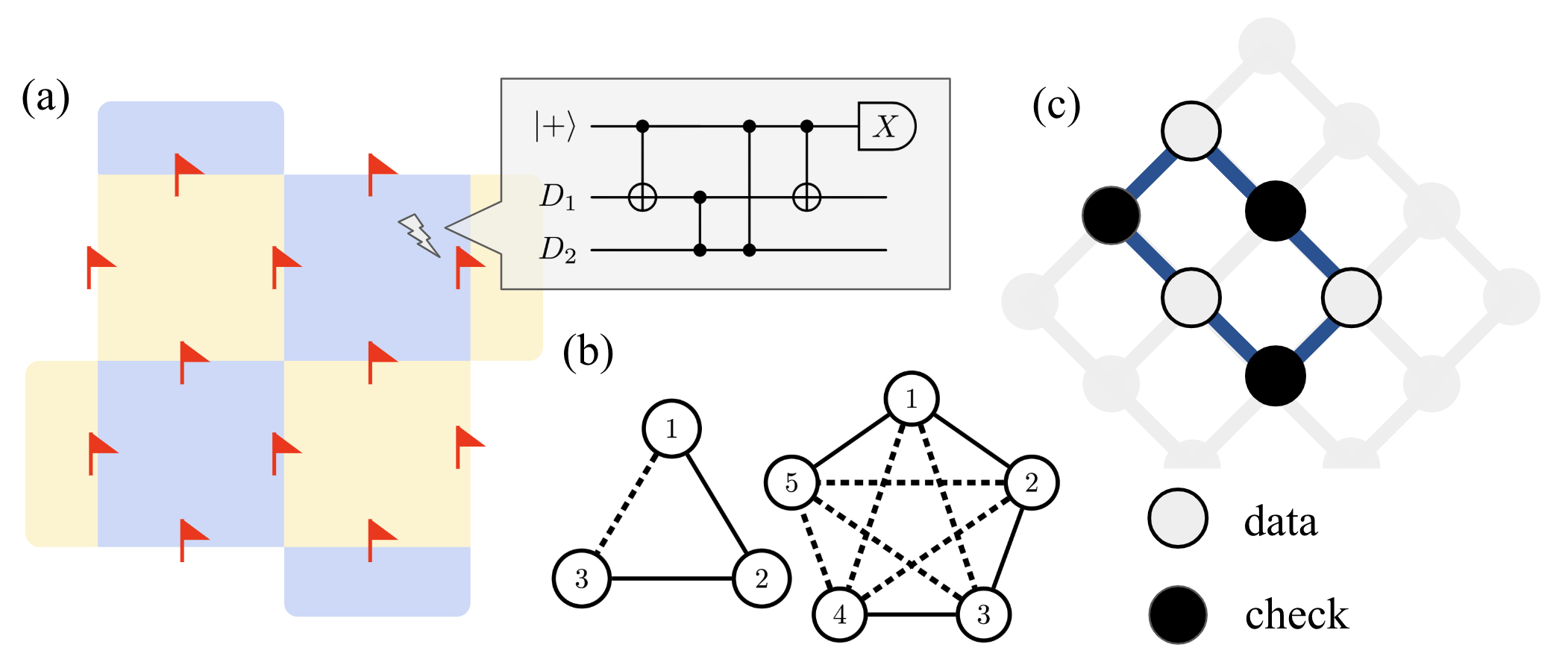}
\caption{Crosstalk-robust quantum error correction circuit engineering. 
Crosstalk-flagged surface code design (a) shows where the flag qubits need to be placed 
and how to implement $ZZ$-flagged fault-tolerant CZ gates between data and ancilla qubits, 
using an additional qubit that is frequently measured and reset in the $X$-basis. 
Figure~(b) is a geometric representation for data qubits (vertices) and stabilizer checks (edges) 
in repetition codes, illustrating the detection of erroneous edges. 
Figure~(c) shows a hardware mapping that adds an extra stabilizer check for a 3-qubit repetition code, 
where blue lines are necessary couplings, and grey (black) circles are data (check) qubits.}
\label{fig2}
\end{figure*}

From the memory and stability experiments, we observe that the most detrimental crosstalk is between data and ancilla qubits, which impacts both memory experiment by decreasing its threshold value and stability experiments in lowering down the logical error rates for a given number of syndrome rounds. To address this issue caused by crosstalk-induced ancilla flips, we propose two crosstalk-robust syndrome extraction schemes: the first scheme uses the crosstalk-flagged gadget to perform two-qubit gates fault-tolerantly; the second scheme uses redundant stabilizer checks to detect flipped ancilla qubits. The two schemes are both hardware-agnostic. While we demonstrate their usage under surface code and repetition code, we would like to note that they are generalizable to more QEC codes.

\subsection{Flagged Syndrome Measurement}
\begin{figure}[htbp]
\centerline{\includegraphics[width=\columnwidth]{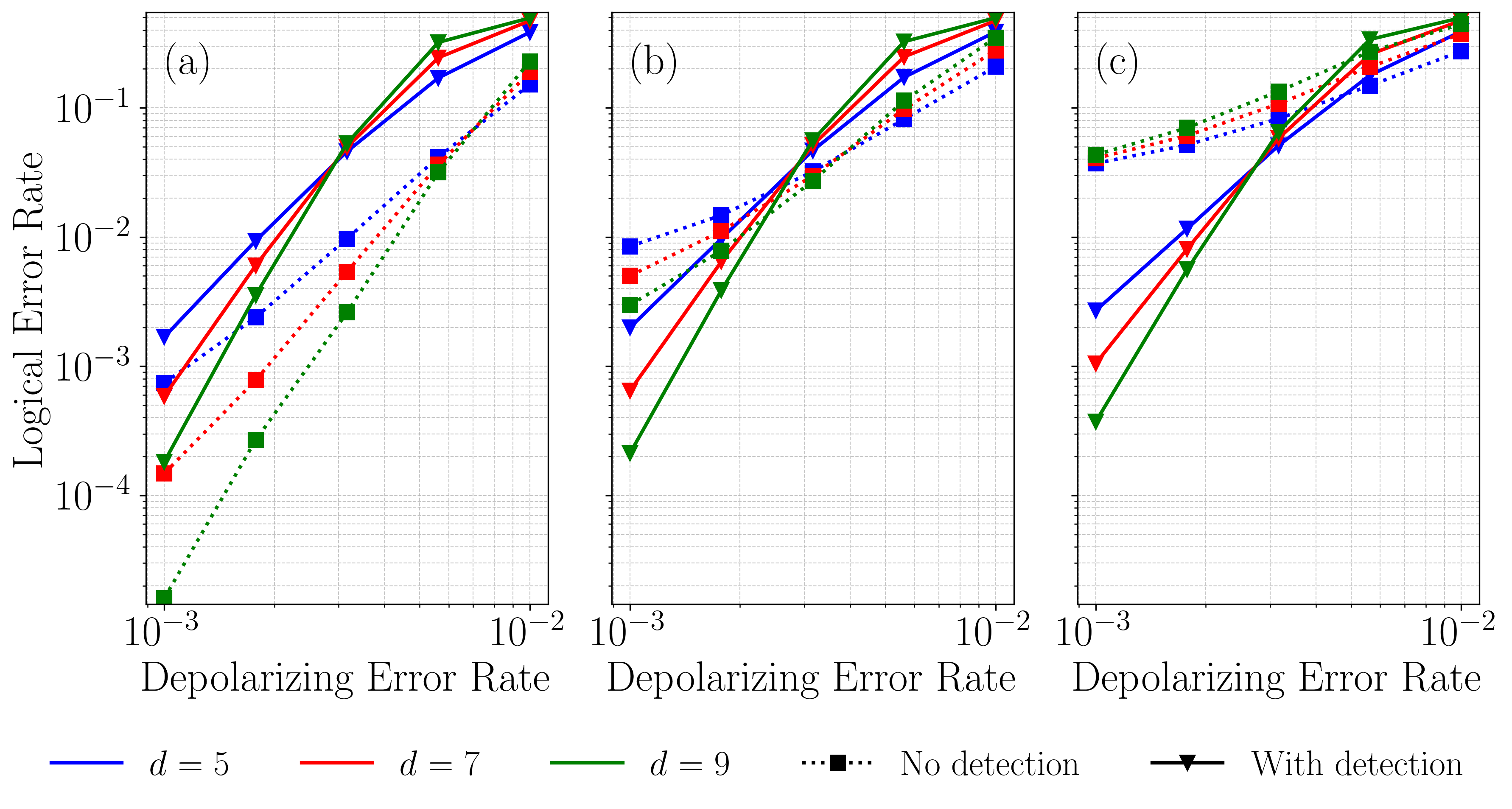}}
\caption{Comparison of logical error rates of rotated surface code memory with or without crosstalk detection through flagged syndrome measurement circuits. The logical error rates for surface code with crosstalk detection (solid lines) is higher than without detection (dashed lines) in low-crosstalk regime of $0.1\%$ (a), is lower than without detection at sub-threshold regime of mid-crosstalk regime of $0.5\%$ (b), and is much lower than without crosstalk in general for high-crosstalk regime of $1\%$ (c). Crosstalk detection are able to stabilize the shifts of threshold caused by higher data-ancilla gate-based crosstalk. In each plot, surface code with distance-$5$(blue), $7$(red), and $9$(green) are shown. }
\label{fig3}
\end{figure}

We first propose to detect crosstalk during syndrome measurements using flag qubits for surface code, inspired from previous flag-qubit designs \cite{Chao2020flag, Chao2018FTfew, Chao2018twoextra}. Our strategy makes use of fault-tolerant gadgets to perform crosstalk-flagged CZ gates for projective measurement, which allows for the detection of at most $1$ $ZZ$-type correlated error between any two qubits. Specifically, in Figure \ref{fig2} (a), between every pair of data qubits there is a flag qubit (red flag), which is coupled to both data and ancilla qubits. The normal CZ gate used from syndrome measurement is replaced with a four-gate gadget for $ZZ$-crosstalk detection. Specifically, the gadget will still perform CZ operation between data and ancilla qubit, but the flag qubit will be measured in $X$-basis constantly where a $-1$ eigenvalue will flag a $ZZ$ correlated error. The samples with detected $ZZ$ errors will be discarded. This setup requires frequent measurement and reset of flag qubit at a rate $4$ times faster than that of ancilla qubits measurements. In this setting, we also need extra flag qubits which scales with distance as $n_{\text{flag}} = 2d^2-2d$, and the number of gates needed for crosstalk-robust syndrome measurement is $4$ time higher than the conventional syndrome measurement circuit (excluding measurements). To quantify the efficacy of using this scheme to perform syndrome measurement, we simulate surface code memory circuit with an effective crosstalk error rate and depolarizing error rate under flagged qubit gadgets scheme. Specifically, the data and ancilla qubit will experience additional depolarizing channel from three additional gates. There will be a residual undetected $ZZ$-crosstalk error (two or four errors) rate that scales as
\begin{equation}
    p_{res} = \frac{6p^2(1-p)^2+p^4}{(1-p)^4+6p^2(1-p)^2+p^4} \approx 6p^2
\end{equation}
with small $p$, the original $ZZ$ crosstalk error rate. Fig. \ref{fig3} shows the logical error rate comparison between memory circuit with or without crosstalk detection under a $ZZ$ error rate of $0.1\%$(a), $0.5\%$(b), and $1\%$(c) for rotated surface codes with distances $d_s \in \{5,7,9\}$. We observed that crosstalk-detected circuits can well stabilize the threshold shifts caused by strong crosstalk between data and ancilla qubits. Crosstalk detection lowers down logical error rate under strong crosstalk ($p>0.5\%$), particularly in the sub-threshold regime. Also, crosstalk-detected surface code offers a higher effective distance $d_{\textrm{eff}}$ (slope) than normal QEC with crosstalk. In the low-crosstalk regime, however, original circuit is more favorable due to less circuit-level depolarizing noise it introduces. 
\subsection{Redundant Stabilizer Checks}
To detect measurement classification errors, we can use redundant stabilizers to check for parity invariance across all possible stabilizer group. This scheme can be useful for QEC codes such as the repetition codes. In a $3$-qubit bit-flip code, for example, checking $s_1=Z_1Z_2$ and $s_2 = Z_2Z_3$ provides sufficient syndrome information to decode the error on one of the three qubits. If there are $1$ or $3$ ancilla errors happen, we can introduce an extra check $s_3 = Z_1Z_3$ to detect: the parity of $s_1s_2$ should equal $s_3$.  For normal $n$-qubit repetition code, we choose $\{P_1P_2, ..., P_{n-1}P_n\}, P\in\{X,Z\}$ as our stabilizer set. By adding one more stabilizer $P_1P_n$, odd number of measurement classification errors can be detected and discarded. This can reduce the overall classification error rate of all stabilizer checks from $p = 1-(1-p)^{n-1}$ to $p_{res} = 1-\frac{2(1-p)^n}{1+(1-2p)^n}$. To realize this on hardware, the mapping is also as natural as connecting the $1D$ chain of repetition code into a loop. Fig.\ref{fig2}(c) shows a $2D$-grid hardware mapping for a $3$-qubit repetition code with an extra stabilizer check. To generalize this scheme for detecting a specific ancilla check, we can formulate this problem as a graph problem: in a graph with $n$ vertices representing  $n$ data qubits, the edges connecting them are the weight-$2$ stabilizer checks. For each stabilizer, to detect its error, each edge needs to be in at least one closed loop. To detect $t$ independent ancilla errors, each one of the $t$ edges needs to be in a distinct closed loop where the total number of erroneous edges in that loop is odd (even number of erroneous edges will be undetectable). Fig.\ref{fig2}(b) shows a geometric representation of constructing redundant stabilizer checks for faulty ancilla qubits.

\section{In-Patch Logical Crosstalk}
\begin{figure}[htbp] \centerline{\includegraphics[width=0.9\columnwidth]{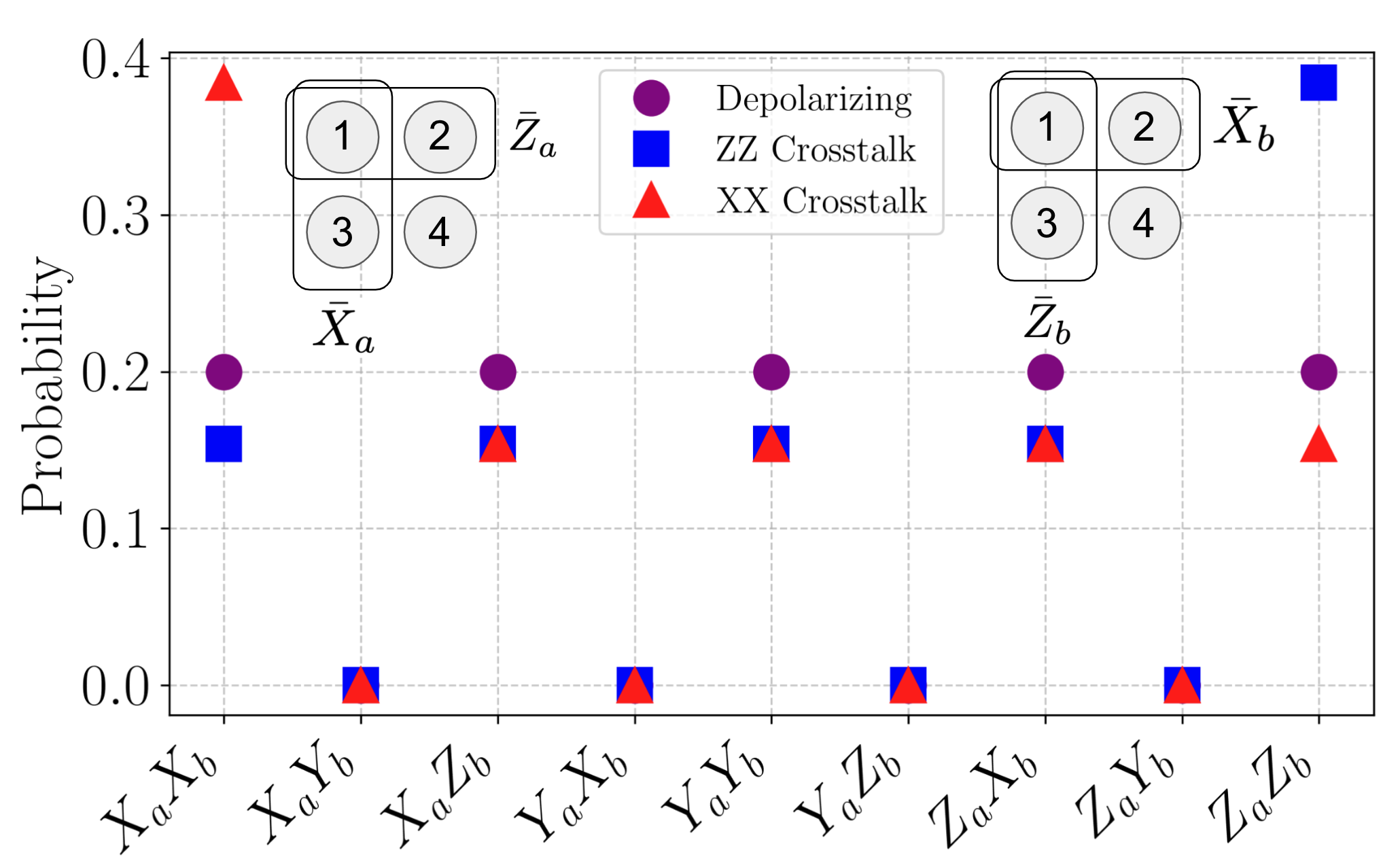}} \caption{The probability of logical crosstalk of different types under three correlated weight-2 error models in a $[[4,2,2]]$ error-detecting code. The types of possible nontrivial logical crosstalk operators are shown on the x-axis. The data points show the probability of logical crosstalk occurring under a 2-qubit depolarizing noise channel (purple), and a 2-qubit depolarizing channel with $XX$ (red) or $ZZ$ (blue) crosstalk. Errors are assumed to occur on all possible pairs of the 4 data qubits with equal probability. The logical operator definitions for this code are illustrated at the top of the plot. All probabilities are normalized.} \label{fig8} \end{figure}

So far, we have studied how physical crosstalk can impact code performance for $[[n, k=1, d]]$ codes. In this section, we discuss the issue of logical crosstalk between logical qubits defined in a $[[n, k>1, d]]$ code block. Starting with the example of the $[[4,2,2]]$ error detection code, the logical Pauli operators can be defined as $\bar{Z}_a = Z_1Z_2$ and $\bar{Z}_b = Z_1Z_3$, as shown in the subplot in Fig.~\ref{fig8}. Hence, if a weight-2 physical crosstalk in the form of $Z_2Z_3$ exists, it is effectively converted to $\bar{Z}_a\bar{Z}_b$, a logical $ZZ$-type crosstalk, due to the overlap of the $Z_1$ operator that supports both logical operators. The same applies to logical $XX$ crosstalk. We calculate and plot the error probability of logical crosstalk for $[[4,2,2]]$ codes in Fig.~\ref{fig8}. In the plot, we show how a 2-qubit depolarizing channel, or a 2-qubit depolarizing channel with either $ZZ$- or $XX$-biased physical crosstalk error, can convert into correlated errors between two logical qubits. We calculate the probabilities assuming that all pairs of the four qubits are equally likely to experience crosstalk. From the plot, we observe that additional biased crosstalk increases the likelihood of logical $XX$- or $ZZ$-type crosstalk, but not other types. Similar issues also occur in other error detection codes, such as the $[[8,3,2]]$ code for weight-2 crosstalk, and more generally in codes where physical crosstalk is of higher weight. Such logically correlated errors cannot be corrected and may affect two-qubit gate error rates.

However, we provide a simple proof that for a $[[n, k \geq 2, d > 2]]$ code, weight-2 physical crosstalk can never be converted into logical crosstalk. This statement can also be generalized to physical crosstalk of arbitrary weight. The statement is as follows:
Let $\mathcal{C}$ be an $[[n,k,d]]$ quantum code. Suppose that physical correlated Pauli errors are at most weight $w_X$. Then, for $\mathcal{C}$, if $d > w_X$, there will be no logical correlated Pauli error of any weight. The following is a proof sketch for the logical crosstalk condition:
\begin{IEEEproof} Assume, for contradiction, that in the given setting there exists a logical correlated Pauli error $E$ in the $[[n,k,d]]$ code. Since the error itself is a logical operator, its weight $w_E$ is at least $d$ ($w_E \geq d$). At the same time, the logical error $E$ is converted from a physical correlated error of weight $w_X = w_E \geq d$. This contradicts the setup of $\mathcal{C}$, where $d > w_X$. Hence, no logical correlated error can occur. \end{IEEEproof}

This shows that if we scale up the code distance, as long as the physical crosstalk weight is less than $d$, it will not be converted into a logical crosstalk error in the code block. In practice, experiments have found that the dominant crosstalk errors are weight-2, with higher-weight errors being sufficiently rare. This suggests that almost any large-distance code would be free from such issues. However, it is important to note that in some scenarios, weight-2 physical crosstalk can propagate and correlate into higher-weight crosstalk, increasing the chance of conversion into logical crosstalk. In particular, in syndrome measurement circuits, a weight-2 crosstalk can propagate through CNOT gates and become, at most, a weight-5 crosstalk involving the four data qubits and one ancilla qubit. To assess and address logical crosstalk, one can perform conversion analysis similar to that in Fig.~\ref{fig8}, which can inform strategies such as crosstalk-aware compiling. This involves intentionally separating key physical qubit pairs (i.e., those where correlated errors can lead to logical crosstalk) when mapping syndrome measurement or algorithm circuits onto physical hardware. This requires further study and development of new compiler techniques or integration into existing ones.
\section{Conclusion}
In this work, we have systematically investigated the impact of crosstalk noise on the performance of QEC codes, focusing on both memory and FTQC experiments. Through extensive numerical simulations, we identified gate-based crosstalk between data and ancilla qubits as the most detrimental type, significantly reducing the error threshold and increasing the time overhead ratio. To address this issue, we proposed two hardware-agnostic strategies: crosstalk-flagged syndrome extraction and redundant stabilizer checks, both of which effectively detect crosstalk-induced errors. We applied both numerical and analytical studies to examine their efficacy. Furthermore, we introduced the concept of logical crosstalk in multi-logical-qubit code blocks, demonstrating that physical crosstalk can propagate to logical errors under certain conditions. We provided analytical criteria to ensure that physical crosstalk does not translate into logical crosstalk, emphasizing the importance of code distance in suppressing such effects.

As an extension of this work, we would like to modify the syndrome decoder to account for correlated errors to improve the scaling of logical error rates. We will also use the insights gained from this study to inform the design of syndrome measurement compilers, particularly for hardware technologies that are susceptible to high crosstalk.

In summary, this work bridges the gap between theoretical QEC studies and practical hardware considerations, offering actionable insights for improving the resilience of quantum error correction in the presence of crosstalk. These results contribute to the broader goal of realizing scalable and fault-tolerant quantum computation.
\section*{Acknowledgment}
We would like to thank Gregory Quiroz, Pei-Kai Tsai, and Kathleen Chang for the useful discussions. This project was supported by the National Science Foundation (under award CCF-2338063). External interest disclosure: YD is a scientific advisor to, and receives consulting fees from Quantum Circuits, Inc.
\bibliographystyle{IEEEtran}
\bibliography{refs}

\end{document}